\documentclass[12pt]{iopart}

\usepackage{float}
\usepackage{units}
\usepackage{graphicx}
\usepackage{hyperref}

\newcommand{\be}{\begin{equation}}
\newcommand{\ee}{\end{equation}}
\newcommand{\bea}{\begin{eqnarray}}
\newcommand{\eea}{\end{eqnarray}}
\newcommand{\degC}{^{\circ}C}
\begin{document}
\title[How many acres of potatoes does a society need?]{How many acres of potatoes does a society need?}
\author{N T Moore}
\footnote{Present address:
Department of Physics, Winona State University, Winona, MN 55987, USA}
\ead{nmoore@winona.edu}
\date{\today}
\begin{abstract}
One of the main difficulties in a class on Sources of Energy and Social Policy is the wide variety of units used by different technologists (BTU's, Barrels of oil, Quads, kWh, etc).  As every student eats, I think some of this confusion can be resolved by starting and grounding the class with a discussion of food and food production.  A general outline for this introduction is provided and two interesting historical cultural examples, Tenochtitlan and the Irish Potato Famine, are provided.  
Science and Social Policy classes are full of bespoke units and involve many different contexts.  Starting the class with a discussion of food energy is a nice way for everyone to start with the same context.  In addition, discussion of Food Energy can lead to interesting historical claims. 
\end{abstract}
\maketitle

\section{Introduction}
When the United States entered World War One one of the problems they faced was logistics.  How much food do you need to ship overseas to Europe to feed a million soldiers?  That early work in nutrition led to the $3000$ Calorie diet many people remember from secondary Health Education class.  A bit about ``Calorie'' (uppercase) vs ``calorie'' (lowercase) units you might remember: $1~Calorie = 1~kilocalorie~(kcal)$, and a dietitian might build a $3000 kcal$ diet for a 20 year old basketball player.  A \textit{calorie} is the amount of energy it takes  to heat a gram of water by a degree Celsius.  There are about $4.2$ Joules in a single calorie, and a Joule occurs all over introductory  physics.  If you need to buy a new home furnace, the sales brochure might advertise that it is capable of delivering $100,000$ BTU's of heat each hour.  What's a BTU? Heat a pound of water by $1^{\circ}F$.  Of course Heat Pumps are far more efficient than simply burning methane or propane, but they consume kilo-watt-hours (kWh) of electricity, not BTU's.  What's a kWh?  Run a $1000$ Watt toaster for an hour and you'll have pulled one kWh off the grid, it will cost you about $\$0.13$ in Minnesota.  If you decide to put solar panels in your backyard, they will probably collect about $10\%$ of the $3.5kWh$ the  the sun delivers to each square meter of your lawn (in Minnesota) each day.  

As the previous paragraph illustrates, there are a frustratingly large number of different units in an ``Energy'' class.  At Winona State, this 3 credit class fulfills a ``Science and Social Policy'' general education requirement and is taken by students from across the university.   Lots of college majors don't require a math class beyond algebra or introductory statistics and the population is largely math-averse. You could jokingly say that one of the main things students learn in the class is unit conversion, but it isn't far off.  Nearly every field finds energy a useful representation, and every profession has their own set of units and terminology most well suited for quick calculation.  Would a medical lab scientist talk about the fractional acre-foot of urine needed test kidney function?  No, but someone in the central valley of California would certainly care about the acre-feet of water necessary to grow almonds!  Does a gas station price their gasoline in dollars per kWh? Given the growing electrification of cars, they might soon.

Everyone eats, maybe not $3000 kcals$ per day, but at least something every day.  When I teach our energy class,
\cite{Energy_textbook,PFFP}, 
I spend a few weeks talking about food energy before all other types.  While food production is not central to climate change and wars over oil, food is essential in a way that diesel and gasoline are not.  Vehicle fuel makes modern life possible, but we could live, unpleasantly, without it.  We can't live without fats and protein.  

\section{Food Energy}

To introduce Food Energy, I ask the students to work through a few questions:

\begin{figure}[h]
\centering
\includegraphics[width=\columnwidth]{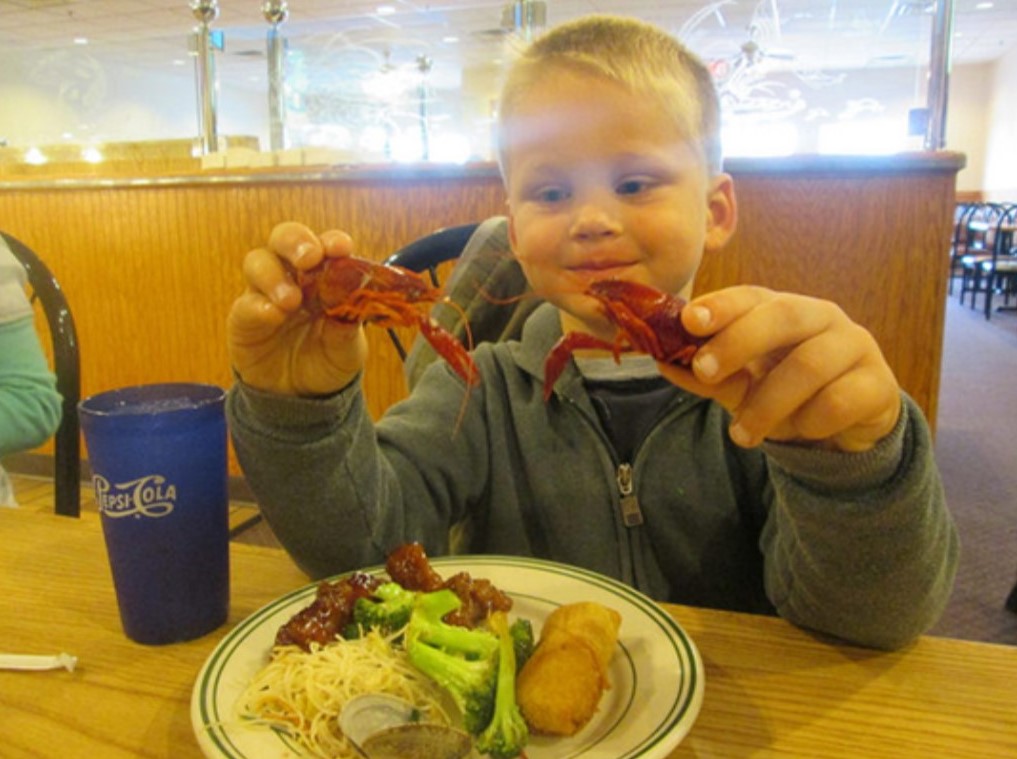}
\caption{
A proto-college-student at Winona's China King Buffet, dreaming about visiting the steam tables every day. 
}
\label{buffet}
\end{figure}

\subsection{Converting food into body heat}
Planning to save money, one college student decides to go to an all-you-can-eat buffet each day at 11am, eg figure \ref{buffet}.  If he brings homework and stretches the meal out for a few hours he can get all $3000kcals$ with only one bill.  Food is fuel for the human body -- could too much fuel make his body feel sick? If his body burned all this food at once, how much warmer would he get? 
Useful information: the student has a mass of $80kg$ and is made mostly of water.  A Calorie heats $1 kg$ of water $1^{\circ}C$. 

Here's a possible answer:
equate food energy with calorimetric heating and assume human bodies have the same heat capacity as water, about $1\frac{kcal}{kg\cdot\degC}$. This allows us to calculate the body's temperature increase.
\bea
3000kcals &=& 80kg\cdot1 \frac{kcal}{kg\cdot \degC}\cdot\Delta T \nonumber \\
\Delta T &\approx& +37.5\degC \nonumber
\eea
Students are normally quite surprised at this number.  Although wildly unrealistic, $\Delta T \approx +6\degC$ is typically fatal, there is a related phenomena of diet-induced thermogenesis\cite{meat_sweats} known informally as ``the meat sweats''. Some students connect this calculation to feeling quite hungry after a cold swim in the pool (a similar effect).  On a larger scale, discussing what's wrong with this estimate is useful.  The main storage mechanism for storing food energy is fat tissue, which the calculation completely ignores.  Infants are generally born with little fat, and an infant sleeping through the night often coincides with the baby developing enough fat tissue to store sufficient kcals to make it though a night without waking up ravenously hungry.  A related follow-up is that if a person is stranded in the wilderness, they should immediately start walking downstream (ie, towards civilization) as they likely won't be able to harvest an amount of kcals equivalent to what they already have stored on their hips and abdomen.\cite{trout}  The contrast of bear hibernation \cite{fat_bear} and songbirds constantly eating through the winter are related connections to investigate.

\subsection{Biophysical Power}
A more realistic question to follow up with relates to the average \textit{power} given off by a person over a day.  
Again, assuming $3000kcal$ is burned over $24 hours$, with useful information: $1 kcal \approx 4200J$ and $1 J/s=1W$.
\be
\frac{3000kcal}{24hours}\cdot\frac{4200J}{1kcal}\cdot\frac{1hour}{3600sec}\approx145W
\ee
Most students still remember $75Watt$ lightbulbs, but given the spread of LED lighting, ``A person's body heat is two $75W$ light bulbs'' will probably only make sense for a few more years.  Desert or cold-weather camping, alone versus with friends, and survival swimming are also examples for students to make sense of this answer.  If you can take advantage of other people's waste body heat, you'll sleep more pleasantly and survive longer in cold water.  

Another application to discuss is that of ``brown fat,'' a sort of biological space heater that humans and other mammals develop in response to cold weather.  This tissue's mitochondria can burn lipids and carbohydrates in a useless proton pumping scheme, which produces metabolic heat \cite{brown_fat_1,brown_fat_2,brown_fat_3,brown_fat_4}.  Most common in rodents and infants, this mechanism can be stimulated by extended exposure to cold temperatures -- the original work was done on lumberjacks in Finland \cite{finland_lumberjacks} .  The idea of a biological space heater that takes a month to turn on and a month to turn off matches the lived experience of college students in Minnesota, who wear down jackets in $4\degC$ weather in November, and beachwear in the same $4\degC$ weather in March.  Additionally, transplants to northern climates often take a few years to ``get used to'' the colder weather up north. It seems just as easy to say that transplants' bodies take a few years to develop the brown fat cells which allow them to be comfortable in cold weather.

One other distinction to emphasize is the difference between power and energy.  A graph of a human body's ``kcal content'' over the course of a day can be a useful illustration.  When sedentary, this graph probably has the slope of $-150W\approx -125 \frac{kcals}{hour}$.  If the $3000kcal$ meal at the buffet takes an hour, this period corresponds to an energy-time slope of 
$+3000\frac{kcal}{hour}\approx +3500W$.  

In medicine, these slopes are effectively equivalent to ``Metabolic Equivalent of Task'' (METS), a common measure in cardiology and exercise physiology.  METS is power normalized by mass, $1METS=1\frac{kcal}{kg\cdot hour}$, and METS levels are available for many different physical activities. \cite{METS}

\subsection{Burning off food energy}
Imagine that after eating a $600kcal$ bacon-maple long-john (donut), you decide to go for a hike to ``work off'' the Calories.  Winona State  is in a river valley bounded by $200m$ tall bluffs.  How high up the bluff would you have to hike to burn off the donut?  
Useful information: human muscle is about $1/3$ efficient, and on Earth's surface, gravitational energy has a slope of about $10~\frac{Joules}{kg\cdot m}$.

\begin{figure}[h]
\centering
\includegraphics[width=\columnwidth]{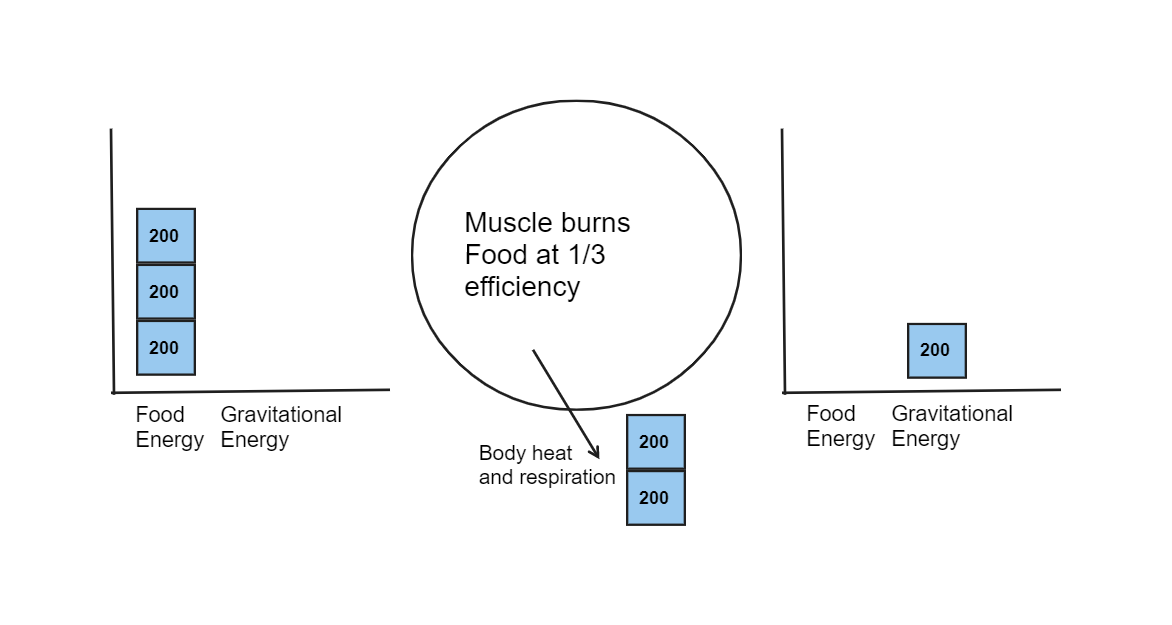}
\caption{An Energy Bar Chart to illustrate the $1/3$ efficient student hiking up a bluff to burn off the morning's donut.  The initial state (left) is the hiker at the bottom of the hill, with donut in stomach.  The final state (right) is the hiker at the top of the bluff with $2/3$ of the energy removed to the atmosphere by sweat and exhalation of warm air. $1/3$ of the donut's energy is stored in elevation.  The system for this diagram includes the earth, the hiker, and the donut.  The system does not include the atmosphere around the hiker.  
}
\label{bar_chart}
\end{figure}

One way to approach this problem is by using Energy Bar Charts \cite{energy_bar_charts} to illustrate how the energy held in food changes form as it is used.  An approximation for this question is shown in figure \ref{bar_chart}.  
In this story, the ``system'' is taken to be the earth, food, and hiker.  The hiker's body is assumed to be $1/3$ efficient, which means one of the food energy blocks of energy is transformed into gravitational energy (elevation) at the end of the hike.  
The other $2$ blocks of energy are transformed into heat and leave the hiker's body, most likely by mechanisms of respiration and sweat evaporation. The purpose of a bar chart like this is to provide a pictorial and mathematical representation of the energy conservation equation given in \ref{eq:bar_chart}.         

\bea
\frac{1}{3}\cdot600kcal\cdot\frac{4200J}{1kcal} 
	&=& 80kg\cdot10\frac{Joules}{kg\cdot m}\cdot height \label{eq:bar_chart}\\
height &\approx&  1000 m
\eea
This estimate is again surprising to students.  Five trips up the bluff to burn off $\$2$ of saturated fat, sugar, and flour!  A nice followup calculation is to imagine a car that can burn a $100kcal$ piece of toast in the engine: from rest, what speed will the toast propel it to? If (again) the engine converts $1/3$ of the energy into motion (kinetic energy), a $1300kg$ Honda Civic will reach a speed of about $15\frac{m}{s}\approx33mph$!  

The point of these energy calculations is not to give students an eating disorder.  Rather, the numbers show food's amazing power. A single slice of toast will bring a car up to the residential speed limit!  A day's food, $3000kcal$, will power you up an $5000m$ mountain peak! The body-work food allows us to do is astonishing, and increases in food production have made modern  comforts, unimaginable $150$ years ago, possible to the point of being taken for granted.  

\clearpage

\subsection{Where does food energy come from?}
One feature of the aught's ``homesteading'' culture \cite{homesteading} is the idea that a person should probably be able to move to the country, eat a lot of peaches, and grow all their own food.  Learning that farming labor is \textit{skilled} labor can be brutal and disheartening. Eating $3000kcals$ each day means planting, weeding, harvesting, and storing more than a million kcals each year \cite{Haspel}.  Where will those Calories come from? Is your backyard enough to homestead in the suburbs \cite{backyard_homestead}?

\begin{figure}[ht!]
\centering
\includegraphics[width=\columnwidth]{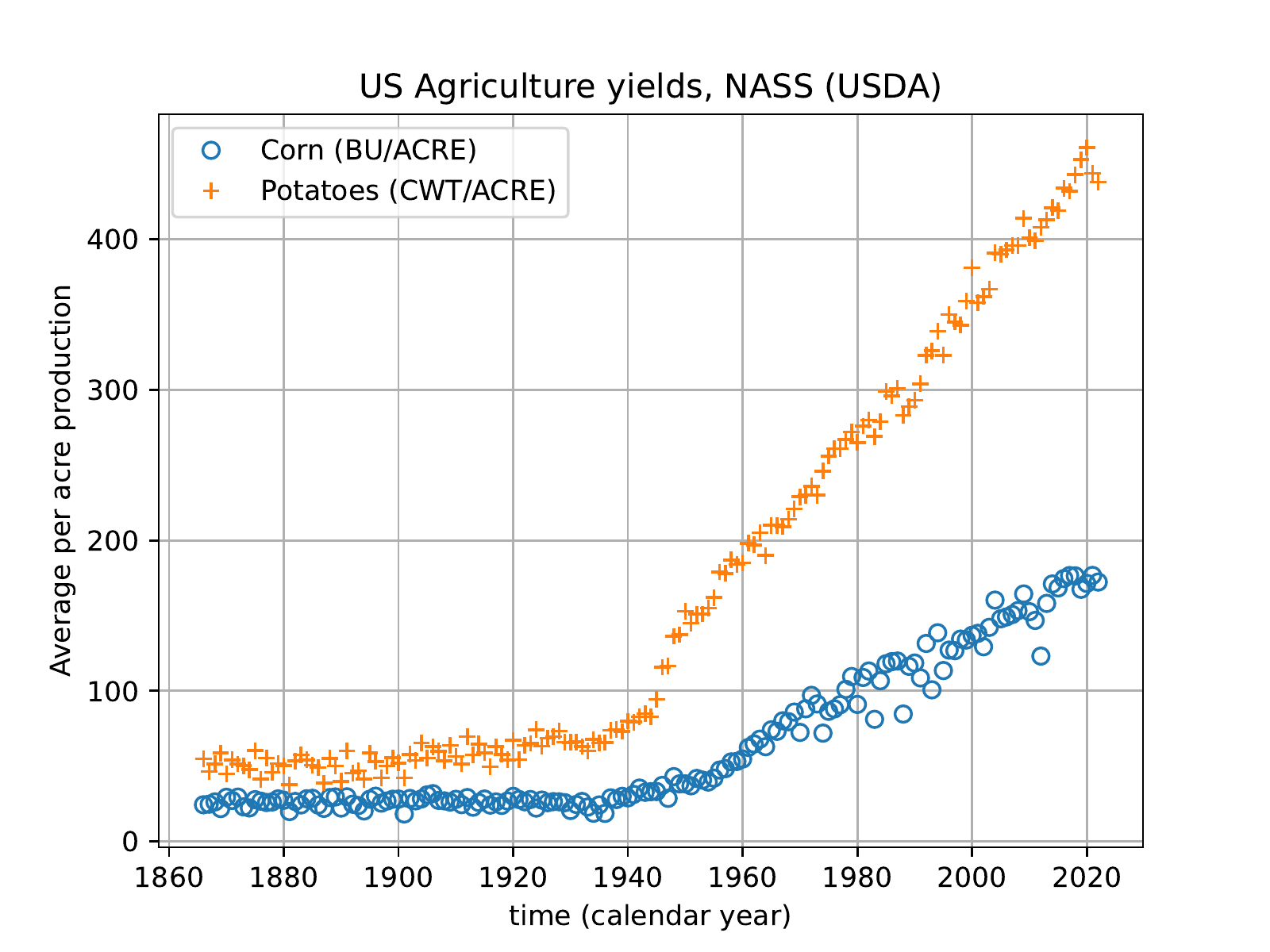}
\caption{
USDA per acre Corn and Potato production figures, plotted over time.  Data is given in harvet units, $56lbs$ bushels per acre for field corn and hundred-weight (CWT) for potatoes.  By mass, corn is about $4.5$ times more calorie dense than potato which results in a nearly equal $kcal/acre$ values for both crops in figure \ref{ag_yields}.
Details on the data source and conversions are given in \ref{how_yield_plot_is_made}.
}
\label{corn_and_potato_yields}
\end{figure}

\begin{figure}[ht!]
\centering
\includegraphics[width=\columnwidth]{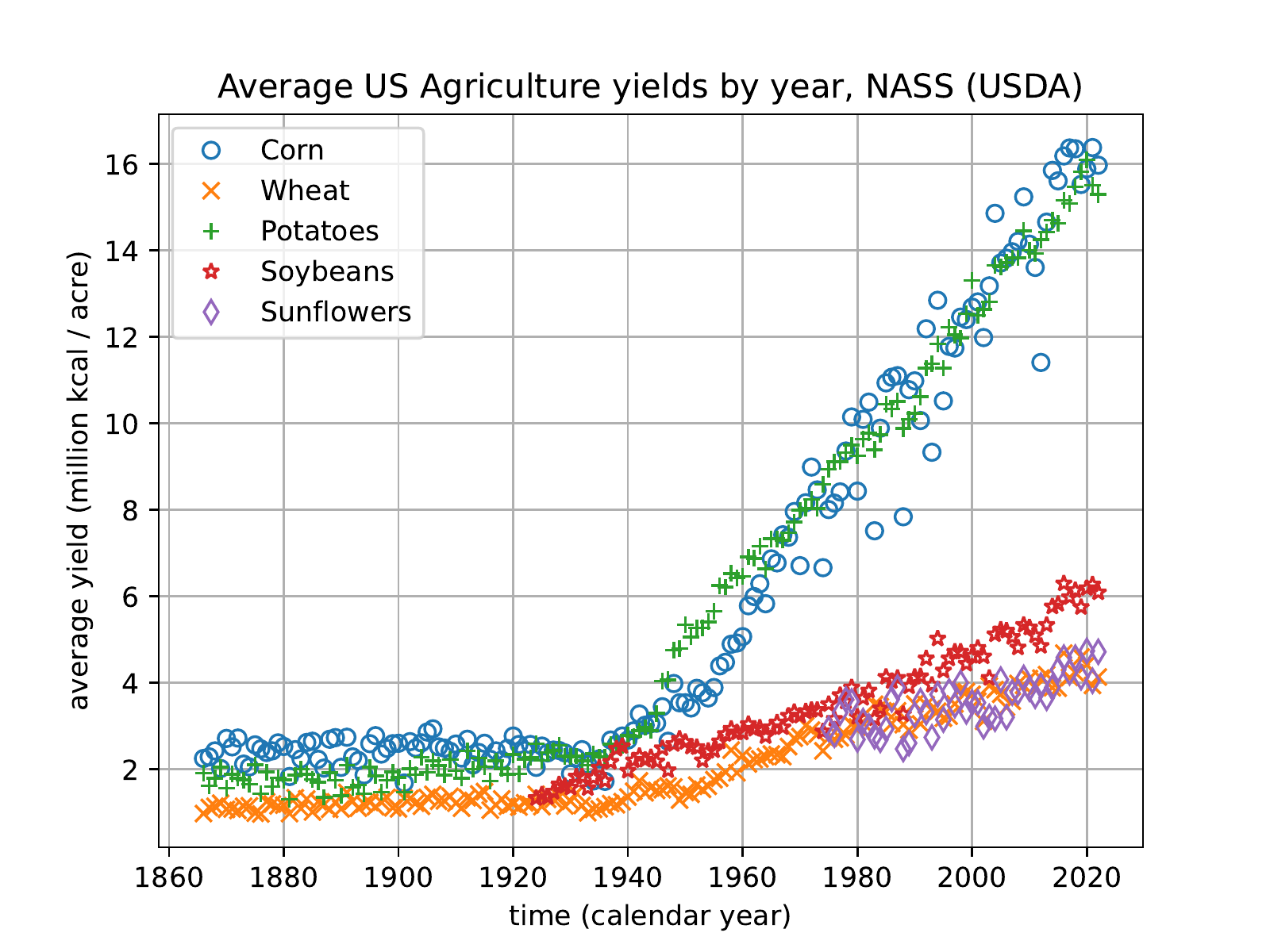}
\caption{
USDA per acre crop production figures, plotted over time.  Production data is scaled by estimated dietary kcal content to show that, over all crops, there has been a dramatic increase in kcal production since about 1940.  
Details of the data source and conversions are given in \ref{how_yield_plot_is_made}.
The idea for this plot came from an online blog, \cite{math_encounters}.  
It would be interesting to know if there are patterns of scaling among vegetable families (grains, legumes, tubers, etc) in the same way that there are family classifications for the minimal energy required for transport \cite{energetic_cost_of_moving}.
}
\label{ag_yields}
\end{figure}

At some point between 1920 and 1950, US chemical manufacturers realized that in the post-war period, they could repurpose processes developed for manufacturing munitions and chemical warfare agents, to produce chemicals that would kill insects and increase the nitrogen levels in the soil. 
As figures \ref{corn_and_potato_yields} and \ref{ag_yields} show, the epoch of ``Better Living Through Chemistry'' produced a dramatic increase in per-acre yields across all comodity food crops, particularly corn and potatoes.  

\begin{figure}[ht!]
\centering
\includegraphics[width=\columnwidth]{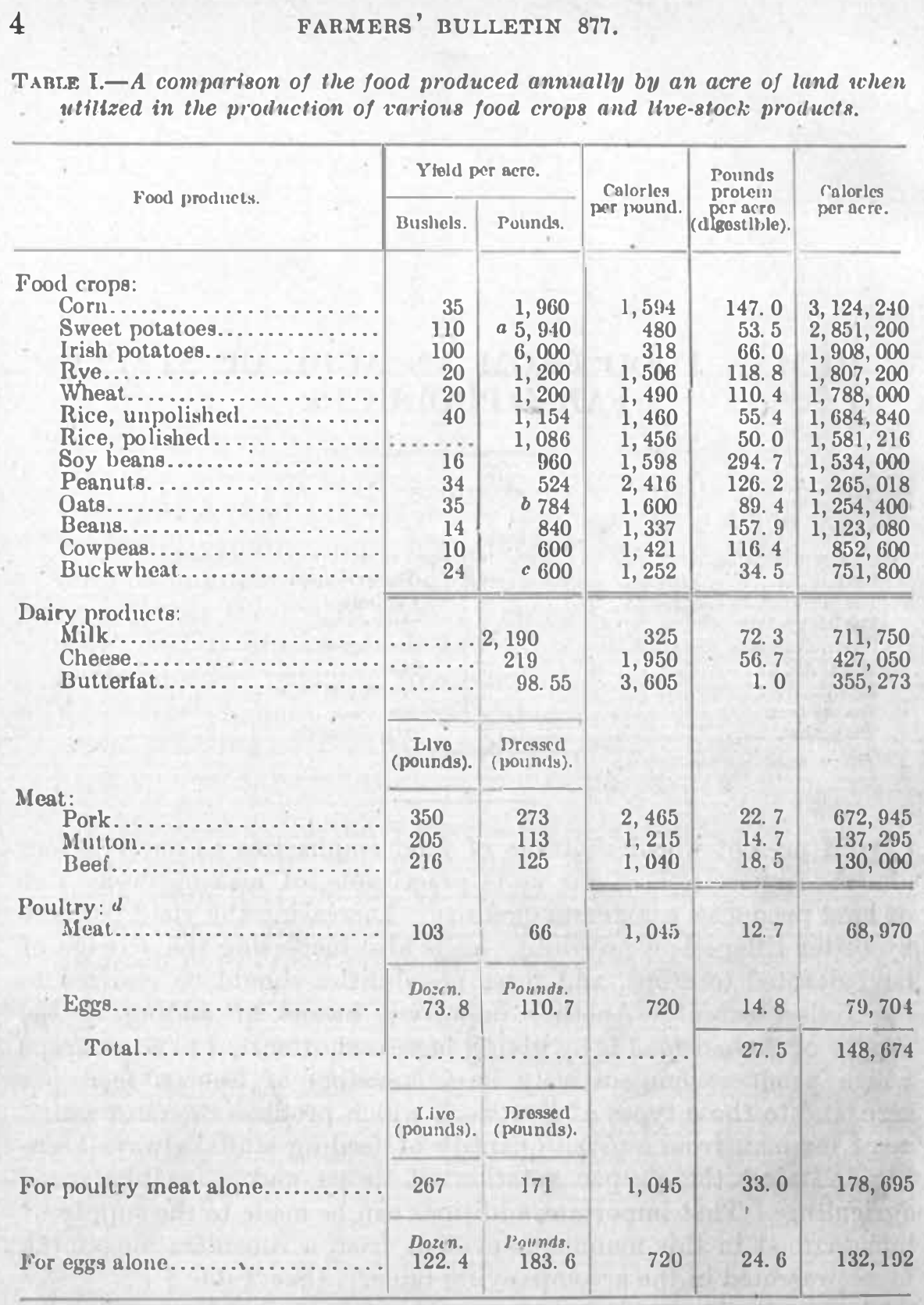}
\caption{
A table from a USDA booklet giving 1917 yields for various farm products.  
}
\label{1917_yields}
\end{figure}

However, if you're discussing backyard Calorie production it isn't reasonable to use modern yield estimates for planning.  ``Roundup Ready'' Corn, Soybean, and Sugar Beet seeds are not available to the public, nobody wants to put on a respirator to apply Atrazine ten feet from the back door,  and the edge effects from deer and insects are much smaller on a $600$ acre field than they are in an community garden allotment.  As mentioned in the introduction, in 1917 the USDA published a pamphlet \cite{USDA_1917_yields_pamphlet} giving detailed Calorie estimates a farmer might expect from a given acre of a crop.  A table from this pamphlet is shown in Figure \ref{1917_yields}.  
The pamphlet data came from pre-war, pre-chemical agriculture, and the yields cited were produced with horses, manure, lime, and large families full of children.  If you want to be self sufficient, these yield numbers are probably a good upper bound on what's realistically possible by a dedicated Luddite.  

So, another question using this data.  If you want to feed your family of four people potatoes, how much land will you need to cultivate?
Here's an estimate: a family of 4 requires $3000kcal/person$ each day\cite{calorie_age}.  If we over-estimate and produce food for the entire year, the family will need about $4.4$ million kcals.
\be
4~people\cdot\frac{3000kcal}{person\cdot day}\cdot\frac{365~days}{year} \approx 4.4 M kcal 
\ee
A brief aside for those bored by the simplistic unit conversion: when I ask students to solve problems like these, one undercurrent of conversation is ``Should I divide by 365 or multiply?''  Particularly with online homework systems, checking your answer for reasonability isn't  typically graded. Asking the students to reason proportionally with units is a skill that can give meaning to numbers. 

From figure \ref{1917_yields} we can estimate $1.9~million~kcals$ per acre of potato production.  Again the students might ask, should I multiple $4.4$ and $1.9$ or should I divide them?  It can be useful in a class discussion to have the students discuss and vote which of the following two forms will give the meaningful answer.
\bea
\frac{4.4 M kcal}{family}\cdot\frac{1 acre}{1.9M kcal}  & \textrm{~~or~~}&
\frac{4.4 M kcal}{family}\cdot\frac{1.9M kcal}{1 acre}
\eea
The choice of operation is difficult to make without seeing the units present, which is again a learning opportunity for the students.

What does the answer of $2.3$ acres mean?  The university's $91m\times49m$ football field has an area of about $1.1$ acres, so you could say that a football field planted in potatoes will probably feed a family through the winter \cite{Deppe}.  Can a person enjoy the benefits of urban living and grow all their own food?  The population density of New Jersey is $1,263~people/mile^2 \approx1.97~people/acre$ and our 4 person family needs $2.3$ acres for their potatoes.  
Unless the social model is one of a country Dacha or an endless suburb with no duplexes or apartment buildings, urban living and food self-sufficiency seem mutually exclusive.

%
%
More emotionally charged conversations can be had about converting the United States to all organic agriculture, which, for corn, typically has a yield penalty of about $20-40bu/acre$ when compared to conventional production.  The 1917 data isn't directly applicable, but it relates. At $180bu/acre$ conventional corn requires $\approx 24~million~acres$ (half of Wisconsin, or all of Indiana) to feed the US population ($350$ million people) corn for a year.  The remainder of the corn belt can be devoted to animal feed, ethanol, and export.  If the corn belt was devoted to producing organic corn at lower yield \cite{organic_corn_yield}, we probably wouldn't starve, but cheap meat and ethanol vehicle fuel would likely disappear.   
%

\clearpage

\section{Example: How big could Tenochtitlan have been?}
The questions described thus far have largely been centered within a physics context.  The paper closes with two more examples that leverage this food energy picture to make historical claims.  The first example relates to the pre-Columbian capital of the Aztec Empire, Tenochtitlan, now known as Mexico City.  Tenochtitlan was built on and around a endorheic lake, Texcoco.  Crops were grown in shallow parts of the lake via chinampas \cite{national_geo}, floating patches of decaying vegetation and soil.  Given the proximity to water and decaying vegetation, these fields were very fertile \cite{HortTech_2019,Chinampas_1964} and some continue to be used in the present day \cite{google_earth}.

Estimates of Tenochtitlan's population in 1500CE vary widely, from 40,000 \cite{40k} to more than 400,000 \cite{400k} inhabitants, comparable in size to Paris at that time. These estimates come from oral and written records and estimates of archaeological building density and land area.   While cannibalism was part of Aztec religious ritual and practice \cite{Aztec_Cannibalism}, the staple Calorie sources for the Aztecs were corn and beans.

Few if any Native American cultures made use of draft animals for food or power before the Columbian Exchange.  This means that the food that fed Tenochtitlan must have been brought to the city center by foot or canoe.  How much land must have been devoted to chinampas to feed the population, or conversely, how many people could be supported by the land within walking or paddling distance from the city center?

A 1964 paper in Scientific American \cite{Chinampas_1964} gives a general outline of the chinampas in the area of Tenochtitlan in 1500CE.  This map seems to be the basis for the similar figure in Wikipedia \cite{chinampas_wikipedia}.  Descriptions of chinampas agriculture indicate that as many as $7$ successive crops could be grown and harvested from the same plot of soil each year, two of which could be maize (corn).  This is truly amazing productivity, given that in the midwest United States corn is normally grown, at most, every other year because of it's extreme nutrient demands on the soil.

There are many ways to approach this estimation problem.  We could assume a Tenochtitlan population of $100,000$ people has a $3000kcal/day$ diet that comes completely from corn.  Assuming that corn's density and nutritional content haven't changed in the $4$ centuries preceding the 1917 data in figure \ref{1917_yields}, we could assume $1lbs$ of corn contains $\approx1594kcal$ of food energy.  
Looking at the map with ImageJ \cite{imageJ}, it seems like the recorded area devoted to chinampas might be about
$16,000~acres$ -- details are given in \ref{appx_imageJ}.
With these assumptions, we could equate the corn energy production from chinampas with the population's yearly food need. Note, in this version of the story, the corn productivity, $P\frac{bu}{acre}$ is treated as an unknown variable.  
\bea
Food~production &=& 16,000acres\cdot \frac{2~corn~crops}{year}\cdot P \frac{bu~of~corn}{acre} \nonumber \\
Population~requires &=& 100,000~people\cdot \frac{3000kcal}{person\cdot day}\cdot\frac{365days}{year}\cdot\frac{1lbs~corn}{1594kcal}\cdot\frac{1bu}{56lbs} \nonumber \\
P \approx 38\frac{bu}{acre} && 
\eea

This crop productivity is in remarkable agreement with the 1917 USDA yields, $35bu/acre$, which seems to validate the assumed $100,000$ person population of Tenochtitlan.  Some references \cite{Chinampas_1964} describe an extensive tribute system that Aztec government required of it's subjects, which certainly would have been necessary to support populations on the upper end of historical estimates \cite{400k}.

\clearpage

\section{Example: Was the Irish Potato Famine a Natural Disaster?}
\begin{figure}[ht!]
\centering
\includegraphics[width=\columnwidth]{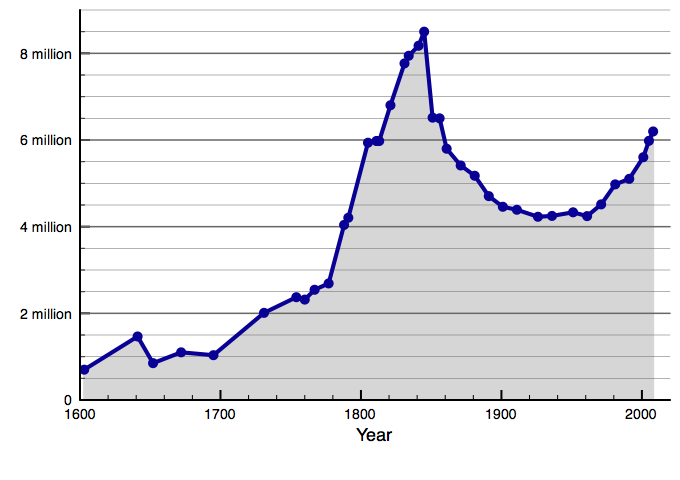}
\caption{
The population of Ireland over time, file from Wikipedia \cite{pop_image}, data sources \cite{pop_sources}. The humble potato, kale, and milk were part of an amazing population boom.  Note that there were two weather-related ``potato'' famines in Ireland, in about 1740 and 1850.  government policy response to the famines could explain the drastic difference in subsequent population following each of the two famines.  The population of Ireland finally re-reached it's 1851 peak in 2021 \cite{Ireland_5M}.  
}
\label{ireland_population}
\end{figure}

In contrast to native cultures of the Americas, Ireland's population boomed with the Columbian Exchange and the introduction of the potato. \cite{potato,little_ice_age}.  Figure \ref{ireland_population} shows that from about 1700 onward there was a dramatic growth in the island's population.  There's never just one reason for historical events, but unlike grains, potatoes thrived in Ireland's cool damp climate and potatoes, kale, and milk form a nutritionally complete diet that greatly reduced hunger-related mortality among the poor working-class in Ireland.  If you look closely at the data in figure \ref{ireland_population} you might believe that there were \textit{two} weather and potato related famines, the most obvious 1845-49 and the second, with much smaller effect on population in 1740-1.  Both famines were precipitated by poor weather, but an important difference is that in 1740, Ireland was a sovereign state but by 1845 the island was effectively an economic colony of the British Empire \cite{little_ice_age}.  

As the story goes, the two main commodity crops in Ireland were potatoes (for humans), and oats, which as horse feed, were something like gasoline in today's economy.  A sovereign government can halt the export of food to feed English horses, which is what happened in 1741 (and 1782). The grain was diverted back as relief to starving people in Ireland, reducing the famine's mortality. However, by 1845 most of Irish farmland was economically controlled by foreign (English) markets, and grain traders typically refused to divert oats (horse feed) as famine relief for the sake of their investment income.

This inflammatory claim, which is certainly a simplified version of history, serves as a useful evaluation example for students. Specifically, in years that the potato crop failed because of weather or late blight, could the amount of oats produced (and exported) have fed the Irish population?  More broadly, was the Great Famine due to weather and disease, natural causes ``we can't do anything about,''  or was the depth of the tragedy a result of political choices?

Some estimates follow:  Ireland's population in 1845 was about $8.5$ million people.  The island has an area of about $84,400km^2$ \cite{IRE_area} and you might estimate that $64\%$ of the land ($54,000km^2$) is arable for agriculture \cite{arable_percentage}.  
It seems reasonable to use the 1917 productivity,  figure \ref{1917_yields},  to make calculations for Ireland in 1845.  Reminder, in 1917, potatoes produced $1.908\times10^6 kcal/acre$ and oats $1.254\times10^6kcal/acre$.
With students, evaluation of the claim could be approached as a series of questions:

How much food does the island need?
\bea
food~needed~per~year &=& 8.5\times10^6~people
	\cdot \frac{3000}{person\cdot day }
	\cdot \frac{365days}{year} \nonumber \\
&\approx& 9.3\times 10^{12} kcals \nonumber
\eea       

How much land area, sown in potatoes, would produce this food?
\bea
9.3\times10^{12}kcals /\left(1.908\times 10^6\frac{kcal}{acre}\right) &=& 4.87\times10^6 acres \nonumber\\
 &\approx& 19,700 km^2 \nonumber
\eea

How much land area, sown in oats, would produce this food?
\bea
9.3\times10^{12}kcals /\left(1.254\times10^6\frac{kcal}{acre}\right) &=& 7.41 \times10^6 acres \nonumber \\
 &\approx& 30,000 km^2 \nonumber
\eea

Summed, $49,700km^2$, these two areas devoted to oats and potatoes are roughly equivalent to the amount of arable land estimated above for Ireland, $54,000km^2$ \cite{arable_percentage}.  What do the numbers mean?  Did there have to be a famine?  If all of the potato crop failed because of late blight, there would likely have been enough oats to feed the population a $2000kcal$ ration of oats with leftover to spare.  
Like the Holodomor or the Great Leap Forward, the numbers suggest that large-scale suffering wasn't  a natural disaster, but rather a human disaster resulting from poor government policy insensitive to the value of human life.

\section{Conclusion}
A class about Energy and Social Policy and the author hasn't mentioned climate change, coal, or solar panels even once!  What is he thinking?  

How many tons of carbon does your car release in a year? How many shiploads of iron oxide will we have to dump into the ocean for phytoplankton to eat up the equivalent about of carbon?  Every question in a class like this is, to at least some extent, informed by numerical calculation and it's pretty arrogant to assume that ``those students'' don't need to (or can't) do the math.  If you're going to have success talking about numerical calculations, you might as well start with examples that everyone can relate to, and everyone eats!  Along the way you might find fascinating historical questions to investigate.

\ack
The work was influenced and improved by discussions with 
Diane Dahle-Koch, 
Larry Moore, 
John Deming, 
Carl Ferkinhoff, 
and Sarah Taber.

\clearpage
\appendix
\section{Creating the historical kcal/acre figure from USDA data}
\label{how_yield_plot_is_made}
The United States Department of Agriculture (USDA) provides historical crop information via the National Agricultureal Statistics Service 
\cite{USDA_NASS} .
Data was downloaded in spreadsheet csv format and then combined and plotted via a Python Jupyter notebook.   

Each crop has its own bespoke units, for example potatoes are sold by hundredweight (CWT) but sugar beets are measured by the ton.  
Every imaginable agricultural product seems to be tracked in the NASS site, for example Maple Syrup production is tracked and given in gallons of syrup per tap! 
Conversion factors used are summarized in Table \ref{conversions}.  
Calorie (kcal) density for each crop was taken from the USDA's Food Data Central
\cite{USDA_FDC}.
Within this database, foods are identified by an FDC ID.  

An example calculation (implemented in the Jupyter notebook) follows for Corn.  
In 2022 the USDA reported an average production of $172.3$ bushels of corn per acre of farmland.  
\be
172.3\frac{bu}{acre}\cdot\frac{56lbs~corn}{bu}\cdot\frac{453.6~grams}{lbs}\cdot\frac{365~kcal}{100~grams} = 15,974,657 \frac{kcal}{acre}
\label{example_calculation}
\ee
Obviously the result is only reasonable to two significant figures!

\begin{table}
\caption{\label{label}
A summary of units and conversions used to create figure \ref{ag_yields} from USDA NASS data.  $1cwt$ is a hundred pounds of potatoes.  
A bushel, $1bu$, is a volume unit of about 35liters and corresponds to about 60lbs of grain. Calorie content per 100 gram (mass) of food is taken from the USDA's ``Food Data Central'' database. 
For context, typical serving sizes are included. 
It isn't clear from any of these resources if lb is pound-force (lbf) or pound-mass (lbm) and so I am treating them as ``grocery store units'' where $1 lbs \approx 453.6 grams$.
}
\begin{indented}
\item[]\begin{tabular}{@{}llllll}
\br
Crop&per acre unit&production unit&kcals per 100gram & typical portion &FDC ID\\
\mr
Corn & bu/acre & $1bu=56lbs$ & 365 & 1 cup is 166g &170288 \\
Potatoes & cwt/acre & $1CWT=100lbs$ & 77 & 0.5 cup is 75g & 170026 \\
Soybeans & bu/acre & $1bu=60lbs$ & 446 & 1 cup is 186g &174270 \\
Sunflowers & lbs/acre & & 584 & 1 cup is 140g & 170562 \\
Wheat & bu/acre & $1bu=60lbs$ & 327 &  1 cup is 192g & 168890 \\
\br
\end{tabular}
\end{indented}
\label{conversions}
\end{table}

\begin{figure}[ht!]
\centering
\includegraphics[width=\columnwidth]{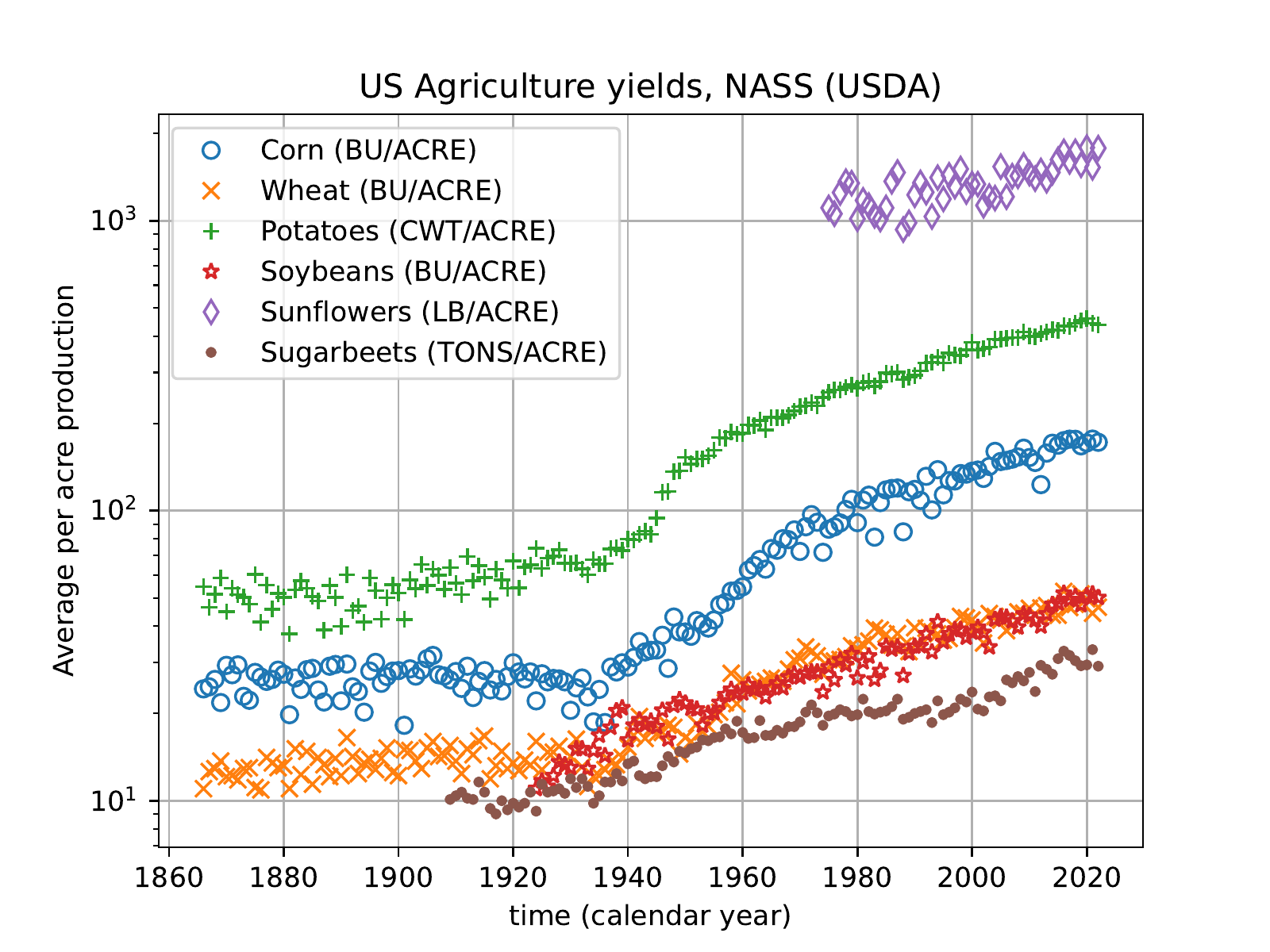}
\caption{
Average USDA per acre yields for a number of commodity crops over time.  This ``raw'' data (in bespoke harvest units) was scaled to produce the data in figure \ref{ag_yields} earlier in the paper. 
}
\label{raw_production_per_acre}
\end{figure}

Raw data from the USDA NASS is plotted in figure \ref{raw_production_per_acre}.  The scaling described in equation \ref{example_calculation} produces figure \ref{ag_yields} earlier in the paper.

\clearpage

\section{Estimating land area devoted to chinampas with ImageJ}
\label{appx_imageJ}

\begin{figure}[ht!]
\centering
\includegraphics[width=\columnwidth]{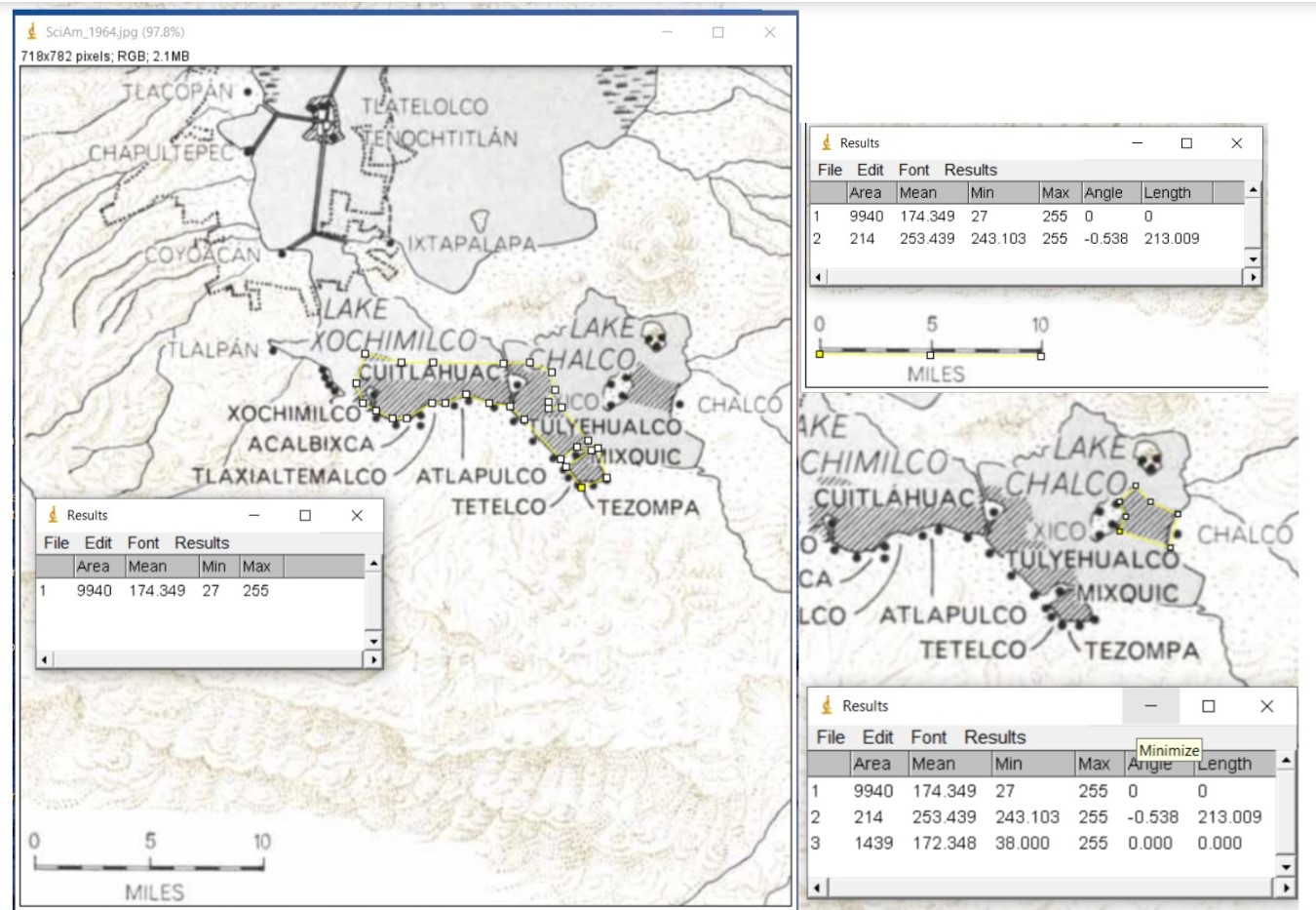}
\caption{
Three screen captures showing chinampa areas and the calibration stick used to convert pixel-squared area into $miles^2$.  The image being analyzed is available in \cite{Chinampas_1964}.
}
\label{imageJ}
\end{figure}

ImageJ is a free software program developed by the National Institutes of Health for photo analysis, \cite{imageJ}.  I used the program to measure a calibration scale in a map and I also used the program to measure the area of two polygons that I drew on the map.  The length and both areas are shown in figure \ref{imageJ}.

Specifically, to find the area of the two large chinampas areas near Tenochtitlan, I took a screenshot from the 1964 paper, \cite{Chinampas_1964}, and saved it in jpg format.  Then, I opened the image in the Windows-Java edition of ImageJ \cite{imageJ}.  The length of the 10 mile distance scale was 213 pixels. The long chinampas area at the south end of the lake was measured with a Polygon selection via the Measure tool to have an area of $9940~pixel^2\approx21.9miles^2$.  The smaller region near Chalco had an area of about $1439~pixel^2\approx3.2miles^2$.  While there were certainly other regions devoted to chimanpas agriculture, the portion visible near the Aztec capital seems to be about $25.1miles^2$ or $16,000acres$.  

\begin{thebibliography}{99}
\bibitem{google_earth}
Chinampas are still visible in sattelite imagery.  See for example latitude = 19.268, longitude -99.087
\bibitem{IRE_area}
Currently, Northern Ireland and the Republic of Ireland are separate countries.  Together, their land area is about $84,400km^2$.
\end{thebibliography}
\end{document}